\documentclass[aps,pra,showpacs,twocolumn,groupedaddress,floatfix]{revtex4-1}

\usepackage{enumerate}
\usepackage{graphicx}
\usepackage{amssymb}
\usepackage{pifont}
\usepackage{amsmath}
\usepackage{multirow}
\usepackage{minibox}
\usepackage{enumerate}
\usepackage{mathtools}
\usepackage{wasysym}

\begin{document}

\def\crta{\vrule height1.41ex depth-1.27ex width0.34em}
\def\dj{d\kern-0.36em\crta}
\def\Crta{\vrule height1ex depth-0.86ex width0.4em}
\def\Dj{D\kern-0.73em\Crta\kern0.33em}
\dimen0=\hsize \dimen1=\hsize \advance\dimen1 by 40pt

\title{It Does Not Follow. Response to ``Yes They Can! \dots''}

\author{Mladen Pavi{\v c}i{\'c}}

\email{mpavicic@irb.hr}

\affiliation{Department of Physics---Nanooptics, 
Faculty of Math. and Natural Sci.~I,
Humboldt University of Berlin, Germany and\\
Center of Excellence for Advanced Materials
and Sensing Devices (CEMS), Photonics and Quantum Optics Unit,
Ru{\dj}er Bo\v skovi\'c Institute, Zagreb, Croatia}

\begin{abstract}
  This a response to ``Yes They Can!\dots'' (a comment on
  \cite{pavicic-nrl-17}) by J.S.~Shaari
  {\it et al.}~\cite{shaari-mancini-pirandola-lucamarini-18}.
  We show that the claims in the comment do not hold up and that
  all the conclusions obtained in \cite{pavicic-nrl-17} are
  correct. In particular, the two considered kinds of two-way
  communication protocols (ping-pong and LM05) under a
  quantum-man-in-the-middle (QMM) attack have neither a critical
  disturbance ($D$), nor a valid privacy amplification (PA)
  procedure, nor an unconditional security proof. However, we
  point out that there is another two-way protocol which does have
  a critical $D$ and a valid PA and which is resistant to a QMM attack.
\end{abstract}



\maketitle

\section{Introduction}
\label{intro}

In Ref.~\cite{pavicic-nrl-17} we considered quantum-man-in-the-middle
(QMM) attacks on two kinds of two-way quantum key
distribution (QKD) protocols: ping-pong one with entangled photons
\cite{bostrom-felbinger-02} and LM05 one with single photons
\cite{lucamarini-05}. In the attacks, an
undetectable eavesdropper (Eve) copies all messages in the
{\it message mode\/} (MM) so as to keep sender's (Bob) qubit in
the quantum memory, sends her own qubits to the encoder (Alice),
receives and decodes Alice's messages, encodes it on kept Bob's
qubits, and sends them back to Bob. We showed that 

\begin{enumerate}[(i)]

\item attacks leave no errors in the MM and hence the standard
methods of establishing security from the literature
\cite{scarani-09} are not available;

\item standard {\it privacy amplification\/} (PA) procedure from
the literature \cite{scarani-09} cannot be executed due to the
absence of a {\it critical value\/} of the {\it disturbance\/}
($D$) in the MM;
  
\item the security proof for LM05 put forward in \cite{lu-cai-11}
does not cover the aforementioned QMM attack and therefore it
cannot be considered to be a proof of {\it unconditional
security\/} of LM05.

\end{enumerate}

The authors of \cite{shaari-mancini-pirandola-lucamarini-18}
claim that the above conclusions obtained in \cite{pavicic-nrl-17}
are erroneous, though. In this paper we show that their claims do
not hold up.

\section{Methods}
\label{methods}

We re-analyze the QMM attacks on two different two-way QKD
protocols: entangled photon ({\it ping-pong\/}) and single photon
(LM05) ones with reference to \cite{pavicic-nrl-17} and
\cite{shaari-mancini-pirandola-lucamarini-18}, in order to show
that the conclusions obtained in \cite{pavicic-nrl-17} are correct
and that the claims against them, put forward in 
\cite{shaari-mancini-pirandola-lucamarini-18}, fail.  

\section{Results and Discussion}
\label{results}

The authors of \cite{shaari-mancini-pirandola-lucamarini-18}
reformulated the above points (i-iii) and we denote those
reformulations as (i)$'$, (ii)$'$, and (iii)$'$, respectively,
below. The points (i)$'$-(iii)$'$ do not faithfully correspond
to (i)-(iii).

In what follows, we show that none of the claims of the authors of
\cite{shaari-mancini-pirandola-lucamarini-18} against (i-iii) holds
up. We analyze their points (i-iii)$'$ one by one and show that
(i-iii) are all correct.

\begin{enumerate}[(i)$'$]
\item attacks which leave no errors in the MM do
not allow for security to be established by standard methods;
\end{enumerate}

The security of a protocol, under individual, or collective, or
general (coherent) attacks, alike, is standardly evaluated via the
critical quantum bit error rate (QBER) by calculating the
secret fraction $r=I_{AB} - I_E$, where $I_E=\min (I_{AE},I_{BE})$,
where $I_{AB}$, $I_{AE}$, and $I_{BE}$, are mutual information in the
MM between Alice and Bob, Alice and Eve, and Bob and Eve, respectively
\cite{scarani-09}. The mutual information serves Alice and Bob to
establish the quantum key distribution (QKD) key in the MM and Eve
to snatch it from the MM.

Now, surprisingly, the authors of
\cite{shaari-mancini-pirandola-lucamarini-18} claim: ``In two-way
QKD $I_E$ is obtained from the CM whereas $I_{AB}$ is estimated from
the MM. Therefore, the CM makes it possible to properly execute the
[privacy amplification] PA whereas the MM enables a proper execution
of the [error correction] EC.'' 

In the {\it control mode\/} (CM) Alice and Bob can only estimate to
which extent Eve was in the line and Eve can find out by which
signals they did so (50\%), but that cannot offer them any
information on the key itself (via PA). On the other hand, there is
no EC, since $I_{AB}=1$.

Contradicting themselves, in the next paragraph
the authors of \cite{shaari-mancini-pirandola-lucamarini-18} admit
that $I_{AB}=1$, that ``there is no error in the MM,''
and that the error rate in the CM is 50\%, but claim that all that ``is
inconsequential'' and write ``Hence, if Eve attacks a fraction $f$ of
the qubits, the resulting secret key rate will be given by
$r=1-f\ge 0$, with equality when $f=1$, i.e., when Eve attacks all
the qubits. Therefore, the protocol is always secure against such an
attack, for any value of $f$.''

However, when Eve attacks all the qubits, she knows all
the messages with certainty, i.e., the whole key. When $f<1$,
Alice and Bob do not have an available procedure to erase the parts
of the key Eve snatched, as shown below. Such a protocol cannot be
considered secure. 

So, \cite{shaari-mancini-pirandola-lucamarini-18}'s claims against (i)
do not hold up. Note that (i)$'$ distorts the meaning of (i) which
claims that ``standard methods,'' i.e., those from the
literature, are not available, not that the security cannot be
achieved at all as stated in (i)$'$. Let us look at the next point. 

\begin{enumerate}[(i)$'$]
\item[(ii)$'$] privacy amplification (PA) cannot be executed due
to the absence of a ``critical value'' similar to BB84's
famous 11\%; 
\end{enumerate}

``PA [is a procedure] aimed at destroying Eve's  knowledge of
the reference raw key\dots The fraction to be removed\dots is $I_E$''
\cite{scarani-09}. The procedure was established in
\cite{bennett-uncond-sec-ieee-95} and refined in
\cite{renner-koening-05}. Eve's knowledge comes exclusively from
MM by the very definition of the mutual information and when
$I_E$ exceeds $I_{AB}$ at the {\it critical value\/} of the 
$D$ (11\% for BB84), Alice and Bob have to abort the transmission. 

In contrast to that, the authors write: ``The absence of errors
in MM simply ensures the unity value for $I_{AB}$, to the legitimate
parties' benefit, and has no bearing in determining the PA rate,
which is a function of $I_E$ estimated from the error rate in CM.
Obviously, differing values for errors in CM would translate into
differing PA rates, hence differing $r$, effectively dismissing the
notion of `critical value'.''

This just does not make any operational sense. All messages Eve
snatched are perfectly correct and there is nothing to be
erased in the PA procedure. On the other hand, due to exponential
attenuation of signals in fibers, Alice and Bob can hardly guess
which messages Eve snatched even when the CM and MM signals are
equally numerous and when $f$ is low. When $f$ is high or close or
equal to 1, this is absolutely impossible because then they erase
the whole key. There is simply no reliable method of identifying
Eve's messages, especially not with the present PA procedure.
So, the considered protocols under the QMM attack is equivalent to
sending a plain text under an unspecified ``protection'' of the CM.  

It helps to dig into the literature. For instance, ``some considered
\dots {\it quantum secure direct communication\/}\break
[two-way protocols] and generated some interest. However, it was
soon recognized that the idea suffers from a major default with
respect to standard QKD. It allows no analogue of privacy amplification:
if an eavesdropper obtains information, it is information on the
message itself and of course cannot be erased.'' \cite{scarani-09}

The authors of  \cite{shaari-mancini-pirandola-lucamarini-18}
further claim that ``a similar argument applies even in the
BB84 protocol, where the amount of PA executed depends on the actual
error rate that has been detected, not on a predetermined critical
value like the 11\% quoted in \cite{pavicic-nrl-17}.''

As stressed above and shown in Fig.~5(a) of \cite{pavicic-nrl-17},
$D=0.11$ is not ``predetermined;'' it corresponds to
$I_{AB}=I_{AE}$. For $D>0.11$ we have $I_{AB}<I_{AE}$, i.e., Eve has
more information than Alice and Bob who therefore cannot possibly
carry out PA beyond $D=0.11$ at which point they are left with no
bits for the key. 

Hence, \cite{shaari-mancini-pirandola-lucamarini-18}'s claims
against (ii) fail, too. Note that (ii)$'$ distorts the meaning of
(ii) which claims that ``standard privacy amplification,'' i.e.,
the one from the literature, cannot be executed, not that the
privacy amplification cannot be executed at all as stated in
(ii)$'$. As for the last point

\begin{enumerate}[(i)$'$]
\item[(iii)$'$] the existing security proofs \cite{lu-cai-11} are
flawed as they do not consider this specific class of attacks,
\end{enumerate}
first of all, nowhere in \cite{pavicic-nrl-17} did we write that
the proofs from \cite{lu-cai-11} are ``flawed.'' We just pointed
out that they do not cover the QMM attack, i.e., that they are not
general enough to provide an unconditional security proof for
the LM05 protocol.

In \cite{shaari-mancini-pirandola-lucamarini-18} we read:
``This is clearly untrue as QMM attack can
be described as a specific case of such proofs. It suffices
to consider Eve's ancilla as a qubit and the unitary
transformation as the well known SWAP gate.''

But, when we look into \cite{lu-cai-11} this does not make
sense. Eq.~(1), III.B.~{\it Eve's attack in Bob-Alice channel\/}
of \cite{lu-cai-11} reads: $U_{BE}|0\rangle_B|E\rangle=
c_{00}|0\rangle_B|E_{00}\rangle+c_{01}|1\rangle_B|E_{01}\rangle$,
etc., where $|0\rangle_B,|1\rangle_B$ are Bob's qubits and
$|E\rangle,|E_{00}\rangle$, $|E_{01}\rangle$ Eve's ancillas.
How can we now consider $U$ as a SWAP gate for her ancillas and what
does it swap for what? Our Eve does not make use of ancillas at all
and does not resend Bob's qubits. So, the security analysis is for
another type of attack, not for the QMM one.

The authors even admit that ``an attack in the backward path is
not made explicit as an extremely pessimistic stand is taken where
Eve is allowed to extract all possible information from the entire
Bob-Eve system without specifying the actual mechanism.''

What does an {\it ``extremely pessimistic stand''\/} mean in the
context of a rigorous proof? Equally so, within a rigorous proof of
a theorem, Eve cannot be {\it ``allowed''\/} to  {\it ``extract all
possible information from the entire Bob-Eve system without
specifying the actual mechanism.''\/} And most importantly,
an analysis of a QMM attack cannot be carried out without
the {\it ``the backward path''\/} being considered.   
If one does not analyse the backward path then one cannot detect
Eve at all because in the QMM Eve just lets all the messages in
both CM and MM through to Alice.

This shows that the proof in \cite{lu-cai-11}
does not cover a QMM attack and, hence, that it is not an
{\it unconditional security proof\/}. Ergo, claims against
(iii) from \cite{shaari-mancini-pirandola-lucamarini-18}
fail, as well.

\section{Conclusions}
\label{concl}

In this paper we analyze the claims put forward by the 
authors of \cite{shaari-mancini-pirandola-lucamarini-18}
that the conclusions obtained in \cite{pavicic-nrl-17} and cited
in Sec.~\ref{intro} as (i), (ii), and (iii), are erroneous.

In Sec.~\ref{results} we show that none of the claims from
\cite{shaari-mancini-pirandola-lucamarini-18} holds up and that
the conclusions (i), (ii), and (iii) about the QMM attacks on
the two-way communication protocols from \cite{pavicic-nrl-17}
are correct.

The most important point of \cite{pavicic-nrl-17} is the following.
It is obvious that when Eve is in the line all the time
(caring out a QMM attack) she has all the messages (without
inducing any error in the MM) and therefore the PA cannot be done.
When Eve is in the line only up to a certain percentage of the
transmission time, then PA might be possible based on the
disturbance ($D$) from the CM but we do not know up to which value
of $D$ (usually called {\em critical\/} $D$) the PA would be valid
under a QMM attack. So, we need an elaborated procedure and
algorithm which would ensure that Eve would possess no
significant number of messages after the PA procedure.
A good example of how to estimate when and how we should
design a PA procedure is given in \cite{kiktenko}.

Having said that, we would like to stress that the two-way direct
communication protocols covered in \cite{pavicic-nrl-17} include
just two best-known ones of their kind: the ping-pong and LM05
protocols (see Sec.~\ref{intro}). These two kinds of two-way
protocols have neither a critical $D$, nor a valid PA for a QMM
attack since no one has come forward with them in the literature
as of yet. The security proof from \cite{lu-cai-11} does not
contain or elaborate on them and therefore cannot be considered
``unconditional'' until one closes this open question.  

Still, a two way communication protocol with a critical $D$,
valid PA, and resistant to a QMM attack, is possible, as shown
in \cite{pavicic-benson-shell-wolters-17}. So, ``quite complex  
two-way QKD security proofs'' which the authors of
\cite{shaari-mancini-pirandola-lucamarini-18} refer to in
their {\it Conclusions\/} might indeed help us all to develop
security analysis of an implementable high capacity (four bits)
two-way protocol.  

\begin{acknowledgments}
Supported by the Croatian Science\break Foundation through project
IP-2014-09-7515 and by the Ministry of Science and Education of
Croatia through the Center of Excellence for Advanced Materials and
Sensing Devices (CEMS) funding as well as by MSE grants
Nos.~KK.01.1.1.01. 0001 and 533-19-15-0022.
\end{acknowledgments}



\bibliographystyle{spmpsci}      

\vfill\eject

\end{document}